%% file: root.tex
\documentclass[letterpaper, 10 pt, conference]{ieeeconf}  

\IEEEoverridecommandlockouts                              

\usepackage{booktabs} 
\usepackage{amsmath}
\usepackage{amssymb}
\usepackage{tabularx}
\usepackage[titlenumbered,ruled,linesnumbered,noend]{algorithm2e}
\SetAlFnt{\footnotesize\sffamily}
\usepackage{makecell}
\usepackage[multiple]{footmisc}
\usepackage{graphicx}
\usepackage{xcolor} 
\usepackage{microtype}

\title{\LARGE \bf
Tradeoff-Focused Contrastive Explanation for MDP Planning
}

\author{Roykrong Sukkerd$^{1}$, Reid Simmons$^{2}$, and David Garlan$^{1}$
\thanks{$^{1}$Institute for Software Research, School of Computer Science,
        Carnegie Mellon University, Pittsburgh, PA 15213, USA
        {\tt\small rsukkerd@cs.cmu.edu, garlan@cs.cmu.edu}}%
\thanks{$^{2}$The Robotics Institute, School of Computer Science,
	Carnegie Mellon University, Pittsburgh, PA 15213, USA
        {\tt\small reids@cs.cmu.edu}}%
}

\begin{document}

\maketitle
\thispagestyle{empty}
\pagestyle{empty}

\begin{abstract}

End-users' trust in automated agents is important as automated decision-making and planning is increasingly used in many aspects of people's lives.
In real-world applications of planning, multiple optimization objectives are often involved. Thus, planning agents' decisions can involve complex tradeoffs among competing objectives. It can be difficult for the end-users to understand why an agent decides on a particular planning solution on the basis of its objective values.
As a result, the users may not know whether the agent is making the right decisions, and may lack trust in it.
In this work, we contribute an approach, based on contrastive explanation, that enables a multi-objective MDP planning agent to explain its decisions in a way that communicates its tradeoff rationale in terms of the domain-level concepts.
We conduct a human subjects experiment to evaluate the effectiveness of our explanation approach in a mobile robot navigation domain. The results show that our approach significantly improves the users' understanding, and confidence in their understanding, of the tradeoff rationale of the planning agent.
\looseness=-1

\end{abstract}


\input{commands}

\input{introduction}
\input{approach}
\input{evaluation}
\input{related-work}
\input{conclusion}

\vspace{.5em}\noindent\textbf{ACKNOWLEDGMENT} This work is supported in part by award N00014172899 from the Office of Naval Research and by the NSA under Award No. H9823018D0008.









\bibliographystyle{IEEEtran}
\bibliography{bib/abbrev-no-acronym,bib/mybib}

\end{document}

%% file: commands.tex
\newcommand{\nb}[2]{
{
 {\color{red}{
 \small\fbox{\bfseries\sffamily\scriptsize#1}
 {\sffamily\small$\triangleright~${\it\sffamily\small #2}$~\triangleleft$}
 }}}
}

\newcommand{\R}{\mathbb{R}}
\newcommand{\Q}{\mathbb{Q}}
\newcommand{\Z}{\mathbb{Z}}
\newcommand{\N}{{\mathbb{N}}} 

\newcommand{\todo}[1]{\nb{TODO}{#1}}
\newcommand{\rk}[2][Roykrong]{\nb{#1}{#2}}
\newcommand{\rs}[2][Reid]{\nb{#1}{#2}}
\newcommand{\dg}[2][David]{\nb{#1}{#2}}

%% file: introduction.tex
\section{Introduction}
\label{sec:introduction}

As automated decision-making and autonomous systems are increasingly used in many aspects of people's lives, it is essential that people have trust in them -- that the systems are making the right decisions and doing the right things.
Gaining such trust requires people to understand, at the appropriate levels of abstractions, why the automated agents or systems made the decisions that they did \cite{gunning2017explainable, fox2017explainable}.

Real-world decision making and planning often involves multiple objectives and uncertainty \cite{white1993survey, marler2004survey, roijers2013survey}. 
Since competing objectives are possible, or even inherent in some domains, an essential part of the reasoning behind multi-objective planning decisions is the tradeoff rationale.
For the end-users to understand why an automated agent makes certain decisions in a multi-objective domain, they would need to be able to recognize when a situation involves tradeoffs, understand how the available actions can impact the different objectives, and understand the specific tradeoffs made by the agent.
This is clearly challenging for end-users when there are a large number of possible, sequential decision choices to consider.
Particularly, it can often be difficult for the users to evaluate the full, long-term consequences of each action choice in terms of the domain-specific concerns of the problem, and to be aware of the interactions among the different concerns.
As a result, the users may not understand why the agent made the decisions that it did. This may potentially undermine the users' trust in the agent.

In this work, we address the challenge of helping end-users understand multi-objective planning rationale by focusing on one of its essential parts: the tradeoff rationale.
We propose an approach for explaining multi-objective Markov decision process (MDP) planning that enables the planning agent to explain its decisions to the user, in a way that communicates the agent's tradeoff reasoning in terms of domain-level concepts.
This work is built upon our preliminary work \cite{sukkerd2018toward},
where we provided an initial framework for contrastive explanations.
Our new contributions include a full formalization of how to compute alternatives for contrastive explanations, and an evaluation of our approach.
The main idea of our approach is to detect and explain whether a situation involves competing objectives, and to explain why the agent selected its planning solution based on the objective values. In instances where there are competing objectives, our approach explains the agent's decisions by explaining how it makes tradeoffs: by contrasting the (expected) consequences of its solution to those of other rational (i.e., non-dominated) alternatives with varying preferences.
\looseness=-1

To evaluate the effectiveness of our explanation approach on end-users, we operationalize the end-users' goal of understanding the agent's decisions to be: to assess whether those decisions are in line with the users' goals, values, and preferences.
In other words, we use a task-oriented evaluation of our explanation approach, where the task of the user is to determine either that: (a)~the agent's planning solution is the best available option with respect to the user's preference for the objectives of the problem domain, or (b)~there exists another available solution that the user would prefer.
The users' ability to make a correct determination serves as a proxy for measuring how well the users understand the agent's planning rationale.
\looseness=-1

\noindent Our contributions are the following:

\paragraph*{(1)} We design an explainable planning approach for multi-objective planning, formalized as a Markov decision process (MDP) with linear scalarization of multiple cost functions. Our approach consists of two parts. (i)~A modeling approach that extends MDP planning representation to ground the cost functions in the domain-specific human-interpretable concepts, to facilitate explanation generation. (ii)~A method for generating \emph{quality-attribute-based contrastive explanations} for a MDP policy, which explain the planning agent's reasoning in the context of other rational decisions that the agent did not make, due to its (a priori) preference.

\paragraph*{(2)} We conduct a human subjects experiment to evaluate the effectiveness of our explanation approach on end-users, with respect to the users' goal of assessing whether the agent makes appropriate decisions. The results show that our approach significantly improves the users' ability and confidence in making correct assessment of the agent's decisions.
\looseness=-1

%% file: approach.tex
\section{Approach}
\label{sec:approach}

The goal of our approach is to enable automatic explanation of multi-objective planning decisions, with the focus on explaining the tradeoff rationale.
In this paper, we refer to the optimization concerns of a given planning problem as \emph{quality attributes (QAs)} \cite{kazman1996scenario} (e.g., execution time, energy consumption, etc.). We focus on Markov decision process (MDP) planning with linear scalarization of multiple objective functions. Our approach supports both the expected total-cost and the long-run average cost criteria of MDP planning.


\subsection{Motivating Example}
\label{sec:example}

\begin{figure}
	\centering
	\includegraphics[trim=0pt 0pt 300pt 0pt, clip, width=1.00\columnwidth]{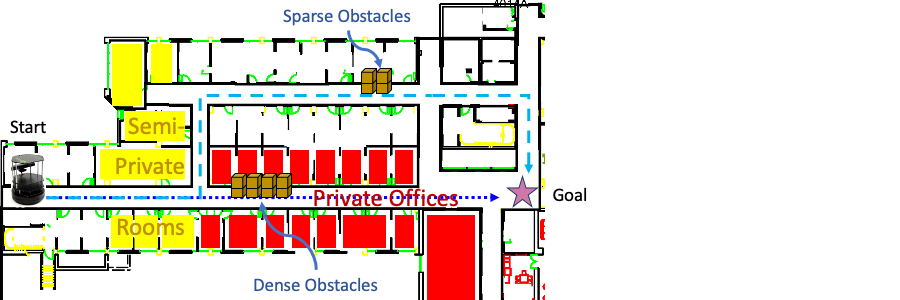}
	\caption{Indoor robot navigation example: The dotted line and the dashed line indicate 2 potential navigation paths. The dotted path has shorter travel time, while the dashed path has lower intrusiveness and lower chance of obstacle collision.}
	\label{fig:mobilerobot}
\end{figure}

We will use the following example to discuss our approach. Figure \ref{fig:mobilerobot} shows a mobile robot whose task is to drive from its current location to a goal location in a building
. The robot has to arrive at the goal as soon as possible, while trying to avoid collisions for its own safety and to avoid driving intrusively through human-occupied areas. The robot has access to the building's map (locations, and connections and distances between them), placement and density of obstacles (sparse or dense obstacles), and the kinds of areas in the environment (public, private, and semi-private areas). The robot can determine its current location, knows its current driving-speed setting, and can detect when it bumps into obstacles. The robot can also navigate between two adjacent locations, and change its driving speed between two settings: full- and half-speed.
\looseness=-1

When the robot drives at its full speed through a path segment that has obstacles, it has some probability of colliding, depending on the density of the obstacles. Furthermore, the intrusiveness of the robot is based on the kinds of areas the robot travels through. The robot is non-intrusive, somewhat intrusive, or very intrusive when it is in a public, semi-private, or private area, respectively.

Figure \ref{fig:mobilerobot} shows an example of a tradeoff reasoning the robot might have to make. Suppose the robot computes the dashed path as its solution. The robot could provide its rationale by showing the alternative dotted path that is more direct, and explaining that while the dashed path takes longer travel time than the direct dotted path, it has lower intrusiveness and lower chance of collision.

Our approach consists of two parts. First is the extended planning representation of MDPs that grounds the cost functions in the quality-attribute semantics of the problem domain, which enables human-interpretable explanation (Section \ref{sec:modeling}). Second is the method for generating quality-attribute-based contrastive explanations for optimal MDP policies (Section \ref{sec:explanation} through Section \ref{sec:soft_constraints}).

\input{approach-modeling}
\input{approach-explanation}
\input{approach-alternatives}

%% file: approach-modeling.tex
\subsection{Modeling Approach}
\label{sec:modeling}

We augment the representation constructs of a factored MDP to preserve the underlying semantics of the quality attributes (QAs) of a problem domain. In this paper, we refer to it as an \emph{explainable} representation. Our approach requires that the designer of an explainable agent specify the planning problem in this representation.

A standard factored MDP is a tuple of the form: $\langle S,A,P,C \rangle$, where $S$ denotes the set of states described via a set of variables $S_1,...,S_n$; $A$ denotes the set of actions; $P$ denotes the probabilistic transition function; and $C$ denotes the cost function, which is a composition of the sub-costs of the QAs. Our approach extends the standard factored MDP constructs to explicitly represents: (i)~the features of state variables and actions that impact the values of the QAs of a planning problem, and (ii)~the analytic models of those QAs.
In addition, our approach requires a mapping from the user-defined types of QAs (e.g., travel time, intrusiveness) 
in the explainable representation to the domain-level vocabulary of the corresponding concepts. This vocabulary will be used to generate explanations.

\subsubsection{Quality-attribute determining factors}
\label{sec:types}

State variables and actions in our explainable representation are typed, and each type defines a set of attributes associated with instances of that type. For instance, the example in Section \ref{sec:example} has a state-variable type {\sf Location}, which has two attributes: {\sf ID} and {\sf Area}, representing the unique ID of a particular location and the type of area that location is in (e.g., public, semi-private, or private area). It also has an action type {\sf MoveTo} (parameterized by a {\sf Location} state variable denoting the destination), which has two attributes: {\sf Distance} and {\sf Obstacle}, representing the distance and obstacle density (e.g., none, sparse obstacles, or dense obstacles) of the path between the source and the target locations of the action. The attribute {\sf Area} of a {\sf Location} state variable has an impact on the intrusiveness of the robot, and the attributes {\sf Distance} and {\sf Obstacle} of a {\sf MoveTo} action have an impact on the travel time and the safety concerns of the robot, respectively.

\subsubsection{Quality-attribute analytic models}
\label{sec:measurements}

A QA $i$ is characterized by a function $\mathit{QA}_i: S \times A \to \mathbb{R}_{\geq 0}$ that calculates the expected quality attribute value for a single decision epoch, that is, the expected value incurs by taking a particular action in a particular state.
Since the state variables and actions are typed, each $\mathit{QA}_i$ is also a function of the attributes associated with the state variables and actions. This maintains explicit measurement models of the QAs in a compact representation.

We consider two types of cost criteria for MDPs: the expected total cost-to-go criterion (when there are absorbing goal states), and the long-run average cost criterion.
For the expected total cost-to-go criterion, the total value of QA $i$ of a policy $\pi$ starting from a state $s$ is characterized by the value function  $J^{\pi}_i(s) = \mathit{QA}_i(s,\pi(s)) + \displaystyle \sum_{s'} P(s'|s,\pi(s)) J^{\pi}_i(s')$.
For the long-run average cost criterion, the average value of QA $i$ of a policy $\pi$ starting from a state $s$ is characterized by the value function $J^{\pi}_i(s) = \displaystyle \lim_{N \to \infty} \frac{1}{N} \mathbb{E} \left\{ \sum_{t=1}^{N} \mathit{QA}_i(s_t,\pi(s_t)) \middle\vert s \right\}$.
For either type of the cost criteria, we consider a multi-objective cost function $C$ of an MDP as a linear scalarization of all $\mathit{QA}_i$ functions: $C(s,a) = \sum_i k_i \mathit{QA}_i(s,a)$, where $k_i > 0$ parameters capture the preference over multiple concerns.

There are different kinds of QAs based on how they are quantified. Thus, they shall be explained differently. Our explainable representation has explicit constructs for defining the following types of QAs:

\paragraph{Counts of Events} QAs of this type are quantified as expected numbers of occurrences of events of interest, where the objectives are to minimize occurrences of ``bad" events. The robot colliding with obstacles is an example of such event, where the safety objective of planning is to minimize the expected number of collisions.

\paragraph{Standard Measurements} QAs of this type can be measured in a standard, scientific way, either directly or by deriving from other measurements. Such QAs are quantified by their magnitudes (e.g., duration, length, rate of change, utilization, etc.), typically measured in real values. The robot's travel time is an example of such measurement. 

\paragraph{Non-Standard Measurements} Some properties do not have associated standard measurements, or their exact standard measurements may not be available to the modelers. Nonetheless, given the domain knowledge, a carefully-defined arbitrary measurement can be used to characterize policies with respect to a particular property of concern. The robot's intrusiveness to the building occupants is an example of such measurement, where penalties can be assigned to indicate the degrees of intrusiveness. We use a set of events $E_m$ to characterize the different degrees of a property $m$ of concern.

An expected sum of 
non-standard measurement values representing an abstract property does not have a well-defined meaning. Unlike for standard measurements, it is not intelligible to communicate a value such as $J^\pi_{intrusive}(s)$ in an explanation. Instead, we propose to communicate the distribution of different non-standard measurement values throughout a policy. To this end, we calculate the expected count of each event in $E_m$ occurring in the policy. We use qualitative terms (e.g., \emph{``not intrusive"}, \emph{``somewhat intrusive"}, and \emph{``very intrusive"}) to describe how often different severity levels of $m$ incur in the policy.


%% file: approach-explanation.tex
\subsection{Policy Explanation}
\label{sec:explanation}

To explain an optimal MDP policy, we propose using \emph{quality-attribute-based contrastive explanations}. For a given planning problem, the explanations: (i)~describe the QA objectives and the expected consequences of the policy in terms of the QA values, and how those values contribute to the overall expected cost of the policy, and (ii)~explains whether (a)~the policy achieves the best possible values for all QAs simultaneously, or (b)~there are competing objectives in this problem instance and explains the tradeoffs made to reconcile them, by contrasting the optimal policy to selected Pareto-optimal alternatives.
In our explanations, we present the MDP policies -- the optimal policy generated by the planner and the alternative policies discussed in the contrastive explanations -- to the user via a custom graphical illustration. A general approach for policy summarization is outside the scope of this work.
\looseness=-1 

\subsubsection*{Policy Consequences}
\label{sec:policy_consequences}

We present an optimal MDP policy, along with the factors that determine its QA values (see Section \ref{sec:types}), via a graphical illustration that employs domain-specific visual components. 
We compute the expected QA consequences of the policy as discussed in Section \ref{sec:measurements}. Such policy evaluation can be done while planning, or afterwards. The corresponding costs of the QA values can be computed in a similar manner.

To generate textual explanation of the planning objectives and the QA values of the solution policy, we use predefined natural-language templates. Recall that different types of QAs should be explained differently. Table \ref{tab:templates} shows examples of explanations of QA objectives and values. The non-italicized terms are originally placeholders in the templates. These placeholders are then filled with either terms from the domain-specific vocabulary provided in a planning problem definition, or QA values computed by our approach.

\begin{table*}[htb]
	\centering
	\begin{tabularx}{\textwidth}{|l|l|X|}
		\hline
		Type of Quality Attribute & Optimization Objective &  Quality Attribute Value \\ \hline
		Count of events & ``\emph{minimize the expected number of} {\sf collisions}" & ``\emph{the expected number of} {\sf collisions} \emph{is} {\sf 0.8}" \\ \hline
		Standard measurement & ``\emph{minimize the expected} {\sf travel time}" & ``\emph{the expected} {\sf travel time} \emph{is} {\sf 10 minutes}" \\ \hline
		Non-standard measurement & ``\emph{minimize the expected} {\sf intrusiveness}" & ``\emph{the robot is expected to be} {\sf non-intrusive} \emph{at} {\sf 5 locations} \emph{and} {\sf somewhat intrusive} \emph{at} {\sf 2 locations}" \\
		\hline
	\end{tabularx}
	\caption{Examples of explanations of quality-attribute objectives and their values.
	\vspace{-3em}}
	\label{tab:templates}
\end{table*}

\subsubsection*{Contrastive Explanations of Tradeoffs}
\label{sec:contrastive_explanation}

Our contrastive explanation approach uses a selected subset of Pareto-optimal policies as rational alternatives to the solution policy, since they reflect the possible tradeoffs among competing optimization objectives. By contrasting the solution policy to each selected Pareto-optimal alternative, we illustrate the gains and losses with respect to the QAs by choosing one policy over the other. This quantitatively explains the QA-tradeoff that the solution policy makes.

We use a predefined natural-language template for generating a verbal explanation: \emph{``I could {\sf [(1) improve these QAs by these amounts]}, by {\sf [(2) carrying out this alternative policy]} instead. However, this would {\sf [(3) worsen these other QAs by these amounts]}. I decided not to do that because {\sf [(4) the improvement in these QAs]} is not worth {\sf [(5) the deterioration in these other QAs]}.}

The statement fragments (1) and (3) contrast the QA values of the solution policy to those of an alternative policy, by describing the gains and the losses quantitatively. (2) describes the alternative policy using a graphical illustration, as discussed in Section \ref{sec:policy_consequences}. (4) and (5) restate the gains and the losses qualitatively, as part of the agent's reasoning to reject the alternative. For instance, the agent rejects an alternative navigation that is 5 minutes shorter than its solution because its expected number of collisions would have been 0.4 higher.
\looseness=-1

On the other hand, if there are QA objectives that are not in conflict with any other objectives, the explanation simply indicates that those QA values of the solution policy are already the best possible values. For instance, the agent's solution is already the least intrusive navigation route.


%% file: approach-alternatives.tex
\subsection{Alternatives Selection for Contrastive Explanations}
\label{sec:alternatives_selection}

To generate contrastive explanations,
our approach obtains alternative policies by finding a Pareto-optimal policy that improves each of the QAs of the solution policy, if one exists. Let $\theta_i$ be a value of the QA $i$ that has $\delta_i$ magnitude improvement from the value of the solution policy (we will discuss more on $\delta_i$ in Section \ref{sec:soft_constraints}). We use $\theta_i$ as an upper-bound constraint on the QA $i$ value for finding a Pareto-optimal, $i$-improving alternative policy.

Specifically, let $\mathcal{M} = \langle S, A, P, C \rangle$ denotes the original MDP problem, where $C(\cdot, \cdot) = \sum_j k_j C_j(\cdot, \cdot)$ is a multi-objective cost function, and each $C_j$ denotes a linear transformation of QA $j$ function for the purpose of normalization. To compute each Pareto-optimal, $i$-improving alternative to the solution policy of $\mathcal{M}$, we solve another MDP $\mathcal{M}^{(i)} = \langle S, A, P, C^{(i)} \rangle$, where $C^{(i)}(\cdot, \cdot) = \sum_{j \neq i} k_j C_j(\cdot, \cdot) + k'_i C_i(\cdot, \cdot)$, where $k'_i$ is relatively small, with the constraint that the expected QA $i$ value of a solution of $\mathcal{M}^{(i)}$ must be $\leq \theta_i$. The specific constraint formulation depends on the cost criterion; as described below.


A standard way of solving constrained MDPs via linear programming \cite{altman1999constrained} will yield randomized stationary policies.
However, in this work, we use only deterministic stationary policies as alternatives to explain the original solution policy, since: (a)~they are more suitable for goal-oriented problems with total-cost criterion, in which decisions would be executed only once, and (b)~they are simpler to describe than randomized policies.

To ensure that our approach produces deterministic alternative policies for the explanations, 
we use the mixed-integer linear program (MILP) formulation in Eqn. \ref{eq:MILP-SSP}, adapted from \cite{DBLP:conf/ijcai/DolgovD05}, for solving a constrained MDP with the expected (undiscounted) total-cost criterion with goals.

\begin{equation}
\label{eq:MILP-SSP}
\begin{array}{rrcl}
 \multicolumn{1}{l}{\displaystyle \min_{x} \displaystyle \sum_s \sum_a x_{sa} C^{(i)}(s,a)} & \multicolumn{3}{l}{\textrm{s.t.}}\\
out(s) - in(s) & = & 0 & \forall s \in S \backslash (G \cup \{s_0\})  \\
 out(s_0) - in(s_0) & = & 1  \\
\displaystyle \sum_{s_{g} \in G} in(s_g) & = & 1 \\
 x_{sa} & \geq & 0 & \forall s \in S, a \in A(s) \\
\displaystyle \sum_{a} \Delta_{sa} & \leq & 1 & \forall s \in S \\
 x_{sa} / X & \leq & \Delta_{sa} & \forall s \in S, a \in A(s) \\
\displaystyle \sum_s \sum_a x_{sa} \mathit{QA}_i(s,a) & \leq & \theta_i \\
\end{array}
\end{equation}
where: (i)~the optimization variables $x_{sa}$ are \emph{occupation measure} of a policy, where $x_{sa}$ represents the expected number of times action $a$ is executed in state $s$; (ii)~$\Delta_{sa} \in \{0,1\}$ are binary variables, (iii)~$X$ is a constant such that $X \geq x_{sa} \, \forall s, a$, which can be computed in polynomial time using the approach such as in \cite{DBLP:conf/ijcai/DolgovD05}; and (iv)~$out(s) = \sum_a x_{s,a}$ and $in(s) = \sum_{s',a} x_{s'a} P(s|s',a)$.
This MILP formulation ensures that for each $s \in S, \, x_{sa} > 0$ for a single $a \in A(s)$.
Once the model is solved, the deterministic solution policy can be recovered as: $\pi(s) = a$ if $x_{sa} > 0$ for all $s \in S$.



For the expected average-cost criterion, we use the MILP formulation in Eqn. \ref{eq:MILP-average-cost} that again ensures deterministic policies. This formulation is applicable to any chain structure of MDPs \cite{DBLP:books/wi/Puterman94}, where: 
(i)~$x_{sa}$ is as defined in Eqn. \ref{eq:MILP-SSP}, and $y_{sa}$ is an additional optimization variables; 
(ii)~$out_{x}(s), out_{y}(s)$ and $in_{x}(s), in_{y}(s)$ are defined in the same way as in Eqn. \ref{eq:MILP-SSP} for $x$ and $y$, respectively; (iii)~$\alpha$ is a distribution of initial states;
(iv)~$\Delta^{(x)}_{sa}$ and $\Delta^{(y)}_{sa}$ are binary variables corresponding to $x$ and $y$, respectively; and
(v)~$X$ and $Y$ are constants such that $X \geq x_{sa}$ and $Y \geq y_{sa} \, \forall s, a$.
We use $X = 1$ and $Y = 1$.

\begin{equation}
\label{eq:MILP-average-cost}
\begin{array}{rrcl}
 \multicolumn{1}{l}{\displaystyle \min_{x,y} \displaystyle \sum_s \sum_a x_{sa} C^{(i)}(s,a)} & \multicolumn{3}{l}{\textrm{s.t.}}\\
 out_x(s) - in_x(s) & = & 0 & \forall s \in S  \\
 out_x(s) + out_y(s) - in_y(s) & = & \alpha_s & \forall s \in S  \\
 x_{sa} & \geq & 0 & \forall s \in S, a \in A(s) \\
 y_{sa} & \geq & 0 & \forall s \in S, a \in A(s) \\
\displaystyle \sum_{a} \Delta^{(x)}_{sa} & \leq & 1 & \forall s \in S \\
 x_{sa} / X & \leq & \Delta^{(x)}_{sa} & \forall s \in S, a \in A(s) \\
\displaystyle \sum_{a} \Delta^{(y)}_{sa} & \leq & 1 & \forall s \in S \\
 y_{sa} / Y & \leq & \Delta^{(y)}_{sa} & \forall s \in S, a \in A(s) \\
\displaystyle \sum_s \sum_a x_{sa} \mathit{QA}_i(s,a) & \leq & \theta_i \\
\end{array}
\end{equation}

Once the model is solved, the corresponding deterministic solution policy can be recovered as:
\[ \pi(s) = 
\begin{cases}
a & \quad \text{if } x_{sa} > 0 \text{ and } s \in S_x \\
a' & \quad \text{if } y_{sa'} > 0 \text{ and } s \in S/S_x,
\end{cases}
\]
where $S_x = \{ s \in S: \sum_a x_{sa} > 0 \}$ is the set of recurrent states, and $S/S_x$ is the set of transient states of the Markov chain generated by $\pi$.
This MILP formulation ensures that for each $s \in S_x, \, x_{sa} > 0$ for a single $a \in A(s)$, and for each $s \in S/S_x, \, y_{sa} > 0$ for a single $a \in A(s)$.


\subsection{Planning with Soft Constraints}
\label{sec:soft_constraints}

As discussed in Section \ref{sec:alternatives_selection}, to determine an upper-bound constraint $\theta_i$ on the QA $i$ value to solve for a Pareto-optimal, $i$-improving alternative policy, we need to determine a minimum magnitude $\delta_i$ of improvement from the value of the solution policy. The simplest option would be to choose a relatively very small $\delta_i$, so that the resulting alternative policy has the next possible (improved) value of QA $i$. However, depending on the planning problem structure, using such approach can yield alternative policies that have QA values that are too similar to those of the solution policy. Contrasting policies that have very similar consequences on the QAs may not be effective in helping the user understand the tradeoffs.

To avoid this issue, one may choose a sufficiently large $\delta_i$, depending on the problem domain, to obtain an alternative policy that is sufficiently different from the solution policy in terms of its QA $i$ value, if one exists. For the purpose of communicating tradeoffs, it is not necessary that the QA $i$ value of the resulting alternative policy is strictly less than $\theta_i$ as long as it is less than that of the solution policy. Thus, it is appropriate to use $\theta_i$ as a soft constraint instead of a hard constraint. Using the method for handling soft constraints in linear programming presented in \cite{DBLP:conf/iccad/CongLZ09}, we can re-formulate the MILP problems in Eqn. \ref{eq:MILP-SSP} and Eqn. \ref{eq:MILP-average-cost} to use $\theta_i$ as a soft constraint by:
\begin{itemize}
	\item Adding a penalty term $\phi_i(v_i)$ to the objective of MILP, where $v_i$ is a variable for the amount of violation of $\theta_i$ and $\phi_i(\cdot)$ is a linear penalty function.
	\item Replacing $\sum_s \sum_a x_{sa} \mathit{QA}_i(s,a) \leq \theta_i$ with the following constraints:
\end{itemize}
 
\begin{equation}
\label{eq:soft-constraint}
\begin{array}{rrclcl}
&\displaystyle \sum_s \sum_a x_{sa} \mathit{QA}_i(s,a) & & \leq & D_i \\
&\displaystyle \sum_s \sum_a x_{sa} \mathit{QA}_i(s,a) & - v_i & \leq & \theta_i \\
& & - v_i & \leq & 0, \\
\end{array}
\end{equation}
where $D_i$ denotes the solution policy's QA $i$ value.

With a linear penalty function, the parameter of the function needs to be tuned to balance minimizing the violation of the soft constraint $\theta_i$ and minimizing the original objective $\sum_s \sum_a x_{sa} C^{(i)}(s,a)$ in Eqn. \ref{eq:MILP-SSP} or Eqn. \ref{eq:MILP-average-cost}. Alternatively, we may use a nonlinear convex penalty function such as quadratic or log barrier function, or any separable convex function, to penalize higher amounts of violation $v_i$ more aggressively relative to minimizing $\sum_s \sum_a x_{sa} C^{(i)}(s,a)$. We use the approach in \cite{DBLP:journals/orl/DAmbrosioLM10} to handle a nonlinear penalty function $\phi_i(\cdot)$ using piecewise linear approximation.

Selecting an appropriate penalty function $\phi_i(\cdot)$ and value for $\theta_i$ can be challenging and is beyond the scope of this paper. Nonetheless, the framework presented here can be further investigated in more specific problem domains and contexts. For instance, $\theta_i$ might come from a user's query in an interactive explanation setting.

%% file: evaluation.tex
\section{Empirical Evaluation}
\label{sec:evaluation}

Recall that the goal of our explanation approach is to help the end-user better understand the agent's reasoning in multi-objective planning, including any tradeoffs it has to make.
As a proxy for measuring such understanding, we measure the user's ability to determine whether the agent's solution is best with respect to a given hypothetical user's preference for the different objectives of the planning domain.
We use this proxy since making the correct determination requires the user to understand the multi-objective reasoning of the agent.
We conducted a user study to evaluate the effectiveness of our explanation approach in improving such ability and confidence of the users in assessing the agent's decisions.

\input{study-design}
\input{analysis-results}
\input{discussion}

%% file: study-design.tex
\subsection{User Study Design}
\label{sec:study}

\paragraph*{Scenario}
We created a user-study scenario in which the user is tasking the mobile robot to deliver a package at some destination in the building, similar to the example in \ref{sec:example}, where the robot has to plan a navigation that minimizes the travel time, expected collisions, and intrusiveness. The user has a particular prioritization of those concerns; however, they do not know a priori whether the robot's planning objective function aligns with their priorities. Once the robot has computed its optimal navigation plan, it will present the plan, as well as the estimated travel time, expected collisions, and intrusiveness of the plan, to the user.

\paragraph*{Task}
The user's task is to determine whether the robot's proposed navigation plan is the best available option according to their priorities.
In the study, to have the ground truth of the preference alignment between the participant (i.e., the user) and the robot, we give a predefined hypothetical preference, in the form of a cost profile, to the participant and instruct them to apply that preference when evaluating the robot's proposed navigation plan. The cost profile consists of a dollar amount per unit of each of the QAs in this domain (e.g., \$1 per 1 second of travel time).

\subsubsection{Questions Design}
\label{sec:questions-design}

Each participant is given a \emph{cost profile} to use as their hypothetical preference, and a series of three \emph{navigation-planning scenarios} plus one additional validating scenario (see Section \ref{sec:answers-validation}). Each scenario consists of: (a)~a building map, and (b)~a proposed navigation plan from the robot. All building maps in all scenarios have the same topological structure, but each map is unique in its random placement of obstacles and random locations of public, semi-private, and private areas. For half of the scenarios, the robot's proposed navigation plan for each scenario is the optimal plan according to the participant's preference (i.e., the robot's plan is aligned with the user's preference). We refer to these as ``preference-aligned" scenarios. For the rest of the scenarios, the robot's proposed navigation plan is computed from a randomly sampled objective function that is misaligned with the participant's preference. We refer to these as ``preference-misaligned" scenarios.
\looseness=-1

Presented with one scenario at a time, the participant is asked to: (1)~indicate whether they think the robot's proposed navigation plan is the best option according to their given cost profile, and (2)~rate their confidence in their answer on a 5-point Likert scale. We refer to this as a \emph{question item}.

In this study, we use a total of 16 different cost profiles, and each cost profile is paired with three different navigation-planning scenarios as described above. Thus, we have a total of 48 unique question items for the participants.

\subsubsection{Validating Participants' Answers}
\label{sec:answers-validation}

We do the following to validate that a participant understands the task correctly. First, we ask the participant to provide the total cost of the robot's proposed navigation plan for each scenario. This is to check whether they understand how to apply a given cost profile to evaluate a plan. The participants are allowed to use a calculator. Second, we embed an ``easy problem" as an extra scenario for each participant, where the robot's proposed navigation plan for that scenario is either the only feasible undominated plan, or a severely suboptimal plan. We exclude the data from any participant who fails to provide reasonable answers for the total cost calculation and the ``easy problem".

\subsection{Experiment Design}
\label{sec:experiment-design}

We recruited 106 participants on the Amazon Mechanical Turk platform with the following standard criteria for selecting legitimate MTurk participants for human subjects research: the participants must be located in the United States and Canada, have completed over 50,000 HITs, have over 97\% approval rating; and have passed our qualification test designed to verify that the participants understand how to apply a given cost profile to evaluate plans. We used a between-subject study design, where we randomly assigned each participant to either the control group or the treatment group. The participants in both groups received the same set of question items, in the same order.
\looseness=-1

Using the validation criteria discussed in Section \ref{sec:answers-validation}, we eliminated 7 participants who gave invalid answers. We collected valid data from the remaining 49 and 50 participants in the control group and the treatment group, respectively.

\paragraph*{Control Group} For each question, participants in the control group are given: 
(a)~the visualization of the robot's proposed navigation plan on the map, and 
(b)~the estimated travel time, expected collision, and intrusiveness of the robot's plan.
\looseness=-1

\paragraph*{Treatment Group} For each question, participants in the treatment group are given the same information as the control group, plus the quality-attribute-based contrastive explanations. The number of alternatives presented in the contrastive explanations is between 1-3 depending on whether the robot's proposed navigation plan can be improved with respect to each objective individually.

We measure the following dependent variables:

\paragraph*{Correctness} The binary correctness of each participant, for each question, in determining whether the robot's proposed navigation plan is the best option with respect to their (given) preference.
\looseness=-1

\paragraph*{Confidence} The confidence level of each participant, for each question, in their answer, ranging from 0 (not confident at all) to 4 (completely confident).

\paragraph*{Reliable Confidence Score} Combining the correctness and confidence measures, we can quantify how well the participants can assess the agent's decisions. We assign a score ranging from $-4$ to $+4$ to each answer, where higher magnitude indicates higher confidence and negative value indicates incorrect answer. The intuition is, we reward higher confidence when correct, and penalize higher confidence when incorrect.
\looseness=-1

We have two central hypotheses in this study:

\noindent \textbf{H1}: Participants who receive the explanations are more likely to correctly determine whether the robot's proposed plan is in line with their (given) preference.

\noindent \textbf{H2}: Participants who receive the explanations have higher confidence in their determination.

%% file: analysis-results.tex
\subsection{Analysis and Results}
\label{sec:analysis}

We collected 297 valid data points from 99 participants. Since each participant provided multiple (three) data points, and different question items (48 unique questions) were used in the study, there may be random effects from the individual participants (e.g., some participants may be better at the task than others) and from the questions (e.g., some questions may be more difficult to answer than others). Therefore, in our statistical analyses, we accounted for those random effects using mixed-effects models \cite{mcculloch2005generalized}. 

\input{results_table}

Table \ref{tab:results} shows the regression analysis results. First column shows the mixed-effects logistic regression result for the \textit{Correctness} binary outcome variable. Second and third columns show the linear mixed-effects regression results for the \textit{Confidence} and the  \textit{Reliable Confidence Score} (or  \textit{Score} for short) outcome variables, respectively.
For all regression analyses, the predictor variables are: (i)~binary flag \textit{is given explanations} indicating whether the participants are given explanations (i.e., whether they are in the treatment group), and (ii)~binary flag \textit{is preference-misaligned scenario} indicating whether the participants are given a ``preference-misaligned" scenario, as opposed to a ``preference-aligned" scenario.
There is no statistically significant interaction between the two predictor variables on either the correctness or confidence. Thus, we do not include the interaction between the predictor variables in our final regression models.

All regression analyses account for two random effects: the random intercepts for individual participants and different question items. We use the random intercept-only models in our final regression models, as the other possible mixed-effects models produce similar results (i.e., are not statistically significantly different).

\subsubsection{Participants' Correctness}
\label{sec:correctness}

To test hypothesis H1, we use mixed-effects logistic regression to find the correlation between the experimental conditions and the correctness of the participants' answers, while accounting for the random intercepts for individual participants and different question items.
\looseness=-1

\paragraph*{Result} Our explanation has a significant positive effect on the participants' correctness. The odds of the participants in the treatment group being correct is on average $e^{1.34} \approx 3.8$ times higher than that of the participants in the control group, with 95\% confidence interval $[2.03, 7.12]$. (Fixed-effect logistic regression coefficient estimate = $1.34$; standard error = $0.32$.) \textbf{H1 is supported}.

The type of scenarios has a significant effect on the participants' correctness, although at a smaller magnitude than the positive effect of explanation. The odds of the participants being correct (regardless of which group they are in) is on average $e^{-1.01} \approx 0.36$ times lower in the ``preference-misaligned" scenario than in the ``preference-aligned" scenario, with 95\% confidence interval $[0.19, 0.70]$. (Fixed-effect logistic regression coefficient estimate = $-1.01$; standard error = $0.33$.)

\subsubsection{Participants' Confidence}
\label{sec:confidence}

To test hypothesis H2, we use linear mixed-effects regression to find the correlation between the experimental conditions and the confidence levels of the participants, while accounting for the random intercepts for individual participants and different question items.

\paragraph*{Result} Our explanation has a significant positive effect on the participants' confidence. The confidence level of the participants in the treatment group is on average $0.42$ higher than that of the participants in the control group, with 95\% confidence interval $[0.09, 0.74]$. (Fixed-effect linear regression coefficient estimate = $0.42$; standard error = $0.17$.) \textbf{H2 is supported}. The effect size is medium: Westfall et al.'s \cite{westfall2014statistical} $d = 0.43$, analogous to Cohen's $d$. 

We do not observe a statistically significant effect of the type of scenarios on the participants' confidence ($p = 0.89$).

\subsubsection{Participants' Reliable Confidence Scores}
\label{sec:score}

Similar to Section \ref{sec:confidence}, we use linear mixed-effects regression to find the correlation between the experimental conditions and the scores, while accounting for the random intercepts for individual participants and different question items.

\paragraph*{Result} Our explanation has a significant positive effect on the participants' scores. The score of the participants in the treatment group is on average $1.72$ higher than that of the participants in the control group, with 95\% confidence interval $[1.03, 2.42]$. (Fixed-effect linear regression coefficient estimate = $1.72$; standard error = $0.36$.) The effect size is medium: Westfall et al.'s $d = 0.62$. 

The type of scenarios has a significant effect of the participants' scores, although at a smaller magnitude than the positive effect of explanation. The score of the participants (regardless of which group they are in) is on average $1.10$ lower in the ``preference-misaligned" scenario than in the ``preference-aligned" scenario, with 95\% confidence interval $[-1.81, -0.38]$. (Fixed-effect linear regression coefficient estimate = $-1.10$; standard error = $0.37$.) The effect size is medium: Westfall et al.'s $d = 0.40$. 

%% file: results_table.tex
\begin{table}[!htbp] \centering   
\begin{tabular}{@{\extracolsep{5pt}}lccc} 
\\[-1.8ex]\hline 
\hline \\[-1.8ex] 
 & \multicolumn{3}{c}{\textit{Dependent variable:}} \\ 
\cline{2-4} 
\\[-1.8ex] & Correctness & Confidence & Score \\ 
\hline \\[-1.8ex] 
 is given explanations & 1.34$^{***}$ & 0.42$^{**}$ & 1.72$^{***}$ \\ 
  & (0.32) & (0.17) & (0.36) \\[0.5em]
 is preference-misaligned & $-$1.01$^{***}$ & $-$0.01 & $-$1.10$^{***}$ \\ 
 scenario & (0.33) & (0.09) & (0.37) \\[0.5em]
 (Intercept) & 0.75$^{***}$ & 2.83$^{***}$ & 0.86$^{***}$ \\ 
  & (0.27) & (0.12) & (0.32) \\ 
\hline \\[-1.8ex]
\textbf{Random Effects} \\
\hline \\[-1.8ex]
\# of Participants & 99 & 99 & 99 \\
Participants Variance & 0.25 & 0.54 & 1.06 \\[0.5em]
\# of Question Items & 48 & 48 & 48 \\
Question Items Variance & 0.29 & 0.00 & 0.45 \\[0.5em]
Residual Variance & & 0.42 & 6.22 \\
\hline \\[-1.8ex]
Observations & 297 & 297 & 297 \\ 
Marginal $R^{2}$ & 0.16 & 0.04 & 0.12 \\
Conditional $R^{2}$ & 0.27 & 0.58 & 0.29 \\
\hline 
\hline \\[-1.8ex] 
\textit{Note:}  & \multicolumn{3}{r}{$^{*}$p$<$0.1; $^{**}$p$<$0.05; $^{***}$p$<$0.01} \\ 
\end{tabular} 
\caption{User Study Results}
\label{tab:results}
\vspace{-2em}
\end{table} 

%% file: discussion.tex
\subsection{Discussion}
\label{sec:discussion}

The results from our user study show that our explanation approach significantly improves the users' ability to correctly determine whether the agent's planning solution is best with respect to their preferences, as well as their confidence in the determination. The users who are given explanations are, on average, $3.8$ times more likely to be correct than those who are not. The explanations provide a moderate improvement in the users' confidence -- with a medium effect size $d = 0.43$.

The results also show that, when the agent's objective function is misaligned with the user's preference, it is more difficult for the user to recognize that the agent's planning solution is not the best option: on average, the users are $0.36$ times less likely to be correct in this type of scenarios. Although, it does not appear that such scenarios significantly affect the users' confidence in their answers.

\subsubsection*{Implication on Real-World Explainable Agency}
Aside from serving as a proxy for measuring the user's understanding of the agent's rationale in multi-objective planning, we argue that the user's ability to assess whether the agent's decisions are best according to their own preference is useful in real-world applications. Potential misalignment between the agent's objective function and the user's preference is inevitable, due to changing preferences over time and contexts, and other limitations of preference learning \cite{leike2018scalable, frnkranz_et_al:DR:2014:4550}.
Our results show that the users have a particularly low probability
of correctly assessing the agent's decisions \emph{when facing value misalignment and given no explanations}.
Thus, we argue that there is a need to improve the users' assessment capability to allow them to properly calibrate their trust in the agent's decisions.
\looseness=-1

To this end, as our results demonstrate significant positive effects of our explanation approach on the users' assessment capability and confidence, we recommend that quality-attribute-based contrastive explanation is a promising approach for explainable agency in real-world applications.

%% file: related-work.tex
\section{Related Work}
\label{sec:related-work}

There are numerous works in the field of explainable AI (XAI) \cite{gunning2017explainable}, but the area that is closely related to our work is explainable agency.
\cite{DBLP:conf/atal/AnjomshoaeNCF19} provides a systematic literature review of works on explainable agency for robots and intelligent agents.
\cite{DBLP:journals/aamas/RosenfeldR19} presents a taxonomy of explainability and a framework designed to enable comparison and evaluation of explainability in human–agent systems.\looseness=-1

Many works in human-robot interaction focus on enabling robotic agents to communicate their goals and objectives to humans.
For instance, \cite{DBLP:journals/arobots/HuangHAD19} contributes an approach to enable a robot to communicate its objective function to humans by selecting scenarios to demonstrate its most informative behaviors.
Similarly, \cite{DBLP:conf/ro-man/LiSASR17} contributes an approach to automatically generate a robot's trajectory demonstrations with particular critical points to improve human observers' abilities to understand and generalize the robot’s state preferences.
Their works share a similar goal with ours, which is to communicate robots' preferences, but the focuses of their contributions are rather in specific domains.
Other works such as \cite{labreuche2011general, DBLP:conf/ijcai/LabreucheF18} propose approaches to explain multi-criteria decision making in non-sequential contexts, but a user-study validation is lacking.\looseness=-1

Many other explainable agent approaches on explaining why agents make certain decisions have been proposed.
Some recent works use argumentation-based approaches for explainable decision making. For instance, \cite{DBLP:conf/aaai/CyrasLMT19} provides argumentative and natural language explanations for scheduling of why a schedule is feasible, efficient or satisfying fixed user decisions.
\cite{DBLP:conf/atal/ZengFMLCO18} presents context-based, explainable decision making via Decision Graphs with Context (DGC).
More closely related to our work are \cite{DBLP:conf/ijcai/0002DSJNICFB19, juozapaitis2019explainable} which use reward decomposition for explaining the decisions of reinforcement learning agents. The approach decomposes rewards into sums of semantically meaningful reward types, so that actions can be compared in terms of tradeoffs among the types.
These works share a similar feature of explanation with ours in terms of illustrating the agent's tradeoffs. However, their work focuses on explaining the tradeoffs of each individual action, while our work differs in that we focus on explaining the tradeoffs of an entire policy via identifying some potential alternative policies.\looseness=-1

Contrastive explanations have been suggested for explainable AI \cite{DBLP:journals/ai/Miller19}, and have been widely used in the field.
Some examples of recent works include \cite{DBLP:conf/ijcai/SreedharanSK18} which provides explanations consisting of information that may be absent in the user's abstract model of the robot's task and show why the foil doesn't apply in the true situation; and \cite{DBLP:conf/ijcai/KimMSAS19} generates explanations for how two sets of plans differ in behavior by inferring LTL specifications that describe temporal differences between two sets of plan traces.
Our work differs in the use of contrastive explanation in that we contrast alternative solutions on the basis of their objective values, in order to explain the tradeoff space of the planning objectives.\looseness=-1

%% file: conclusion.tex
\section{Conclusion}
\label{sec:conclusion}

We present an explainable multi-objective planning approach that focuses on explaining the tradeoffs involved in the decision making, and communicating the planning agent's preference for the different competing objectives, to the end-users. Our approach produces a quality-attribute-based contrastive explanation, which explains the reasoning behind selecting a particular planning solution on the basis of its objective values, in the context of other rational alternative solutions that were not selected. The explanations ground the objective values of planning solutions in the human-interpretable concepts in their respective problem domains. Formally, we contribute a modeling approach for explainable multi-objective MDP planning, and a method for determining Pareto-optimal alternative planning solutions to be used in contrastive explanation. Our user study results show that our explanation approach has a significant positive effect on the users' ability and confidence in understanding the tradeoff rationale of the planning agent, and in determining appropriateness of the agent's decisions.